\documentclass[%
 reprint,
 amsmath,amssymb,
 aps,longbibliography
]{revtex4-1}

\usepackage{graphicx}
\usepackage{dcolumn}
\usepackage{bm}

\begin{document}

\preprint{APS/123-QED}

\title{Current-Modulated Magnetoplasmonic Devices}

\author{Mark E. Nowakowski}
  \email{Mark.Nowakowski@ngc.com}
\affiliation{%
Northrup Grumman Corporation, 1212 Winterson Rd., Linthicum Heights, Maryland 21090, USA
}%

\date{\today}

\begin{abstract}
We model the operation and readout sensitivity of two current-modulated magnetoplasmonic devices which exploit spin Hall effect-like behavior as a function of their device and material parameters. In both devices, current pulses are applied to an electrically-isolated stack, containing an active layer (either a metal with large spin orbit coupling or a topological insulator) embedded within a plasmonic metal (Au). The first device, composed of a ferromagnet and the active layer, illustrates a plasmonic readout scheme for detecting magnetic reorientation driven by current-induced spin transfer torques. The plasmonic readout of these current-modulated non-volatile states may facilitate the development of plasmon-based memory or logic devices. The second device, containing only the active layer, explores the magnetoplasmonic readout conditions required to directly measure the spin accumulation in these materials. The estimated thickness-dependent sensitivity agrees with recent experimental magneto-optical Kerr effect observations.
\end{abstract}

\pacs{Valid PACS appear here}
\maketitle



Actively modulating micro and nano-sized plasmonic devices provides avenues for manipulating, controlling, and directing on-chip surface plasmon polaritons (SPPs) which have been exploited for many applications including super resolution imaging [\onlinecite{Jiang18},\onlinecite{Willets17}] and nanoscale sensing [\onlinecite{Bao12}]. Manipulating these SPP properties is ultimately tied to modifying the dielectric environment near the dielectric/metal interface in these devices and many efforts have demonstrated promising control methods, including: electric field modulation in CMOS-like devices [\onlinecite{Dionne09},\onlinecite{Lee14}] and ultrafast laser-induced modulation [\onlinecite{MacDonald09}]. In addition to these examples, the dielectric environment of a plasmonic heterostructure containing an embedded ferromagnet has also been shown to depend on the external magnetic field-driven orientation of the remanent magnetization [\onlinecite{Temnov10}], this result has motivated research in the area of active magnetoplasmonics [\onlinecite{Armelles13}]. Alternatively, many groups have shown that current-induced spin transfer torques (STTs) [\onlinecite{Stiles09}], generated by spin accumulations along the surfaces of metals with large spin orbit coupling (SOC) [\onlinecite{Liu12},\onlinecite{Pai12}] or topological insulators (TIs) [\onlinecite{Mellnik14},\onlinecite{Li14}], can reorient on-chip magnets without requiring external magnetic fields. This effect has led to the development of spin transfer torque magnetic random access memory (STT-MRAM) [\onlinecite{Apalkov16}].

Here we investigate and estimate the magnetoplasmonic response and sensitivity of two heterostructures containing active layers consisting of materials that generate current-induced spin accumulations. The first heterostructure considers a device which fuses an electrically controlled STT-like bilayer (ferromagnet/active layer) with magnetoplasmonic readout to create a non-volatile plasmonic memory-like element. The second heterostructure estimates the magnetoplasmonic sensitivity required to directly detect current-induced spin accumulations within the active layer and compares the expected detection performance to existing optical methods [\onlinecite{Pattabi15},\onlinecite{Stamm17},\onlinecite{Li18}]. Both of these proposed devices require embedded dielectric layers to electrically isolate the active layer. The losses associated with these additional layers within the context of a magnetoplasmonic heterostructure have not been estimated. In this manuscript, we model the depth, dielectric constant, magnetic strength, and spin diffusion dependences of the magnetoplasmonic behavior and evanescent fields for the proposed devices. These results help define realistic bounds for choosing materials compatible for these demonstrations and sets expectations for the signal strength in fabricated devices.

\begin{figure}[b]
\includegraphics{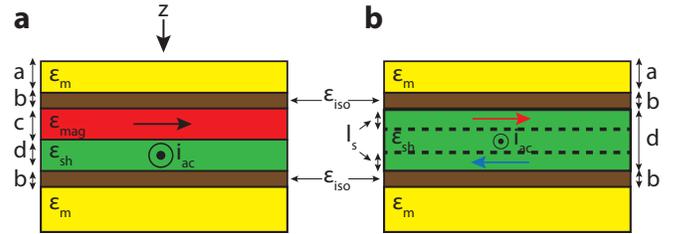}
\caption{\label{fig:epsart1} a,b) Sections of the proposed current-modulated magnetoplasmonic devices. The respective dielectric constants ($\epsilon{_m}$, $\epsilon{_{iso}}$, $\epsilon{_{mag}}$, and $\epsilon{_{sh}}$) and layer thickness ($\it{a}$, $\it{b}$, $\it{c}$, and $\it{d}$) are indicated. In part b, the magnetic layer is removed and the spin diffusion length of the spin Hall layer is $l{_s}$.}
\end{figure}

A cross section for the first proposed heterostructure is shown in Figure 1a. It is a bilayer stack composed of a thin magnet (thickness ${\it{c}}$, dielectric constant ${\epsilon_{mag}}$) directly adjacent to a metal with large SOC or a TI (${\it{d}}$, ${\epsilon_{sh}}$) encased between two thin dielectric layers (${\it{b}}$, ${\epsilon_{iso}}$), surrounded on either side by a metal optimized for plasmon propagation (${\epsilon_{m}}$) where the top layer has a thickness ${\it{a}}$ and the bottom layer is assumed to be large compared to the other layer thicknesses. The second heterostructure shown in Figure 1b is similar to the first stack, however the ferromagnet has been removed. A plasmonic interferometer similar to one used in Ref. [\onlinecite{Temnov10}] is fabricated along the surface of each stack for readout (Supplementary Information S1). We have chosen dielectric constants in each layer assuming the wavelength ${\lambda}$ = 808 nm, however studies have indicated that magnetic SPP modulation is possible over a broad range of wavelengths [\onlinecite{Temnov10},\onlinecite{Becerra10}].

The model for the first proposed heterostructure examines the  magnetoplasmonic influence of bilayer stacks composed of a 5 nm CoFeB magnet and a 10 nm thick layer of either Pt or W encased within two 2 nm MgO dielectric layers [\onlinecite{Liu12},\onlinecite{Pai12},\onlinecite{Li14}]. In these materials the spin Hall effect generates the surface spin accumulation, therefore we generally call this electrically active layer the “spin Hall” layer for brevity. Alternatively, a toplogical insulator, such as Bi${_2}$Se${_3}$, which generates similar surface spin accumulations via spin momentum locking [\onlinecite{Mellnik14},\onlinecite{Li14}] could replace these spin Hall metals, however that case is not considered.

In the first device, the magnetic orientation dependent plasmonic signal level $\lvert\Delta k{_{mp}}\rvert = \lvert k{_x}{^{M=1}}-k{_x}{^{M=-1}}\rvert$ and  SPP propagation length, $ L{_{SPP}}{^{M=\pm1}}$,  are estimated by solving for the dispersion relation of this system, where $M = \pm1$ indicates the antiparallel magnetic orientations and $k{_x}{^{M=\pm1}}$ are the heterostructure propagation constants for each case. Solutions are obtained numerically by solving a system of equations consisting of the wave equation in each material layer and the determinant of a matrix formed from enforcing boundary conditions at each interface (Supplementary Information S2) [\onlinecite{Torrado13}]. Notably, the dielectric tensor of the magnetic layer possesses off-diagonal elements, $\epsilon{_{xz}}$, which are limited to first order. The parity of $\epsilon{_{xz}}$ is magnetic orientation dependent and influences the electric field component of the SPP in the model giving rise to the magnetoplasmonic signal $\lvert\Delta k{_{mp}}\rvert$ (Supplementary Information S2).

\begin{figure}[b]
\includegraphics{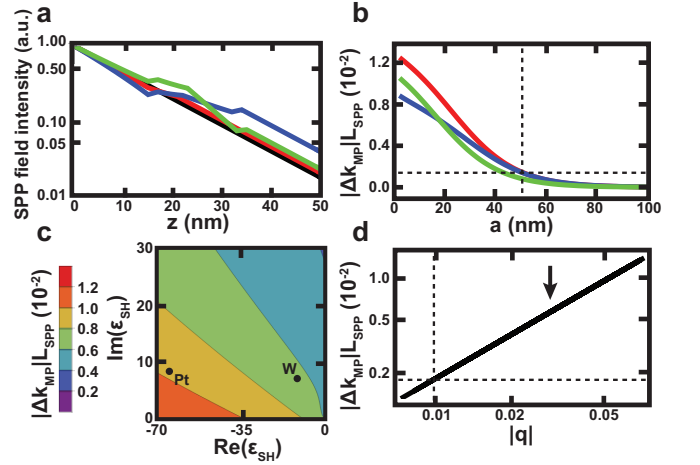}
\caption{\label{fig:epsart2} a) SPP field intensity vs. depth for air/Au (black), Au(15)/Co(6)/Au (red), and Au(15)/MgO(2)/CoFeB(5)/XX(10)/MgO(2)/Au where XX is either Pt (green) or W (blue). b) $\lvert\Delta k{_{mp}}\rvert L{_{SPP}}$ vs. capping Au thickness for Au($\it{a}$)/Co(6)/Au (red) and Au($\it{a}$)/MgO(2)/CoFeB(5)/XX(10)/MgO(2)/Au for Pt (green) and W (blue). Dashed lines indicates detection limit measured in Ref. [\onlinecite{Temnov10}]. c) $\lvert\Delta k{_{mp}}\rvert L{_{SPP}}$ vs. Re($\epsilon{_{sh}}$) and Im($\epsilon{_{sh}}$) for Au(15)/MgO(2)/CoFeB(5)/XX(10)/MgO(2)/Au. Values for $\epsilon{_{Pt}}$ and $\epsilon{_{W}}$ are indicated. d) $\lvert\Delta k{_{mp}}\rvert L{_{SPP}}$ vs. $\lvert\it{q}\rvert$ for Au(15)/MgO(2)/CoFeB(5)/W(10)/MgO(2)/Au. Dashed lines indicate detection limits identified in part b and value used in part a-c is indicated with an arrow.}
\end{figure}

Fig.~\ref{fig:epsart2}a plots the SPP field profile as a function of depth for four structures: a standard air/Au interface (black), a Au(15 nm)/Co(6)/Au geometry similar to Ref. [\onlinecite{Temnov10}] (red), and the proposed geometries, Au(15)/MgO(2)/CoFeB(5)/XX(10)/MgO(2)/Au where XX is either Pt (green) or W (blue). These curves are generated assuming complex optical dielectric constants obtained from the literature at  $\lambda$ = 808 nm for each material: $\epsilon{_{Au}}$ = -24.8+1.5i, $\epsilon{_{Co}}$ = -17.1+24.2i, $\epsilon{_{CoFeB}}$ = -0.33+21.45i, $\epsilon{_{Pt}}$= -66.4+9.52i, $\epsilon{_W}$ = -12.8+8.01i, and  $\epsilon{_{MgO}}$ = 9.8. In the model, the influence of the magnetic layer is $\it{q}= \frac{i \epsilon{_{xz}}}{\epsilon{_{mag}}}$ where $\epsilon{_{mag}}$ is either  $\epsilon{_{Co}}$ or  $\epsilon{_{CoFeB}}$, respectively. For both Co and CoFeB, we assume the value of $\it{q}$ = -0.0345+0.01i [\onlinecite{Temnov10}]. These data indicate that the presence of Pt, W, and the MgO layers do not contribute to a significant deviation of the field profile compared to the air/Au interface, in a manner similar to the Au/Co/Au stack. In fact, the dielectric layers enhance the field strength. From this plot, we conclude that the small perturbations to the SPP environment introduced by the Pt, W, and MgO thin films do not prohibit the proposed device functionality.

The potential effectiveness of the proposed device is explored in Fig.~\ref{fig:epsart2}b which compares the magnetoplasmonic sensitivity of the Ref. [\onlinecite{Temnov10}] stack (red) to the proposed stack with either a 10 nm Pt (green) or W (blue) layer as a function of the capping Au thickness $\it{a}$. The magnetoplasmonic sensitivity is the product$\lvert\Delta k{_{mp}}\rvert L{_{SPP}}$ and is a figure of merit which accounts for both the real and imaginary contributions of $k{_x}{^{M=\pm1}}$  [\onlinecite{Becerra10}]. As $\it{a}$ increases, the active bilayer is pushed farther from the surface reducing its sensitivity. In Ref. [\onlinecite{Temnov10}] a magnetic signal was measured from a layer embedded approximately 50 nm deep into the surface, this point is indicated on Fig.~\ref{fig:epsart2}b and its sensitivity value is approximately 0.002. We consider this value the minimum signal threshold required to detect an analogous current-induced magnetic reorientation. According to the model, the sensitivity of both proposed devices meet this criterion for thicknesses $\it{a} <$ 40 nm in Pt and $\it{a} <$ 50 nm in W. Even though the maximal values in all three proposed devices are less than the Au/Co/Au stack, we expect, based on previous experimental results, that a current-induced magnetic reorientation will be observable using the interferometer geometry.

The spin Hall layer could be replaced with another material, such as a TI [\onlinecite{Mellnik14},\onlinecite{Li14}]. The complex optical dielectric constants for such thin film replacements may not be known. The complex optical dielectric parameter space for the stack: Au(15)/MgO(2)/CoFeB(5)/XX(10)/MgO(2)/Au where XX is the replacement spin Hall layer is explored in Fig.~\ref{fig:epsart2}c. The influence on the magnetoplasmonic sensitivity of the replacement layer XX is studied by varying the real and imaginary optical constants between -70 to 0 and 0 to 30, respectively. We indicate the values for Pt and W we used in Fig.~\ref{fig:epsart2}b on this plot. Using the same sensitivity threshold as before, this plot suggests that a wide variety of complex optical dielectric values are suitable, thus replacement materials for the electrically active layer may also produce robust and detectable current-induced magnetoplasmonic behavior.

Until now, we have assumed a fixed contribution from the off-diagonal magnetic dielectric term $\it{q}$, however the magnitude of this value is material-dependent. In Fig.~\ref{fig:epsart2}d, we plot the magnetoplasmonic sensitivity for a range of $\lvert\it{q}\rvert$, as the real and imagninary components of $\it{q}$ are varied from 0 to 0.05, for the stack: Au(15)/MgO(2)/CoFeB(5)/W(10)/MgO(2)/Au, and indicate the value we have used in our previous plots. Because $\epsilon{_{xz}}$ is taken to first order, we observe a linear dependence between the magnetoplasmonic sensitivity and $\lvert\it{q}\rvert$. This relationship is important to consider within the context of choosing a ferromagnet as a non-volatile memory-like element within a magnetoplasmonic device. While the strength of $\lvert\it{q}\rvert$ may vary between ferromagnets, these results indicate that reasonable sensitivities ($\lvert\Delta k{_{mp}}\rvert L{_{SPP}}>$0.002) may be obtained for values of $\lvert\it{q}\rvert>$0.01, this is approximately three times lower than the value we had previously assumed, and suggests a wide variety of magnets could be selected as the active memory component for this device. By integrating the electrically-isolated bilayer, the magnetoplasmonic modulation of a device consisting of the proposed heterostructure can potentially operate in the GHz regime [\onlinecite{Garello14}]. Exploiting these switching speeds and the non-volatile nature of the proposed heterostructure could provide a pathway towards a plasmon-based STT-MRAM that may be compatible with existing on-chip photonics technologies.

\begin{figure}[b]
\includegraphics{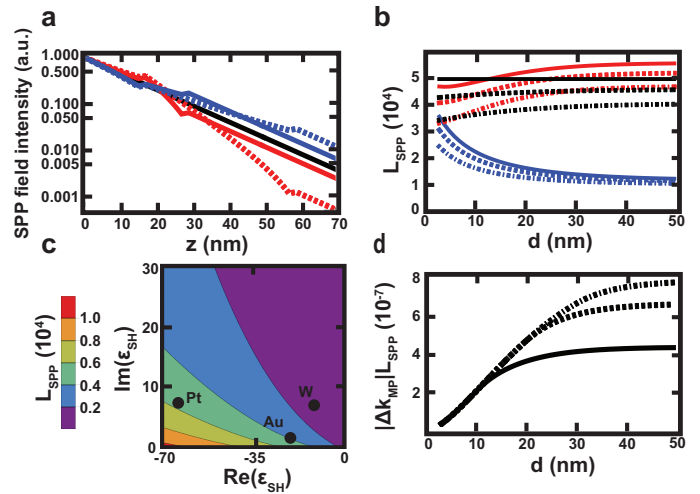}
\caption{\label{fig:epsart3} a) SPP field intensity vs. depth for air/Au (black) and Au(15)/MgO(2)/XX($\it{d}$)/MgO(2)/Au where XX is either Pt (red) or W (blue) for $\it{d}$ = 10 (solid) or 40 nm (dashed). b) $ L{_{SPP}}$ vs. spin Hall and dielectric layer thickness for Au(15)/MgO($\it{b}$)/XX($\it{d}$)/MgO($\it{b}$)/Au where XX is Au (black), Pt (red), and W (blue) for $\it{b}$ = 0 (solid), 2 (dashed), and 5 nm (dot-dashed). c) $L{_{SPP}}$ vs. Re($\epsilon{_{sh}}$) and Im($\epsilon{_{sh}}$) for Au(15)/MgO(2)/XX(40)/MgO(2)/Au. Values for $\epsilon{_{Pt}}$, $\epsilon{_{W}}$, and $\epsilon{_{Au}}$ are indicated. d) $\lvert\Delta k{_{mp}}\rvert L{_{SPP}}$ vs. $\it{d}$ for Au(15)/MgO(2)/Pt($\it{d}$)/MgO(2)/Au for $\it{l}{_s}$ = 5, 10, and 15 nm (solid, dashed, and dot-dashed).}
\end{figure}

The second heterostructure we propose is similar to the first, however the ferromagnetic layer has been removed in order to directly detect the current-induced spin accumulation generated within the active layer; the magnetic signal of this heterostructure, thus, will be reduced. Unlike the previous device, the magnetic signal is only generated when a current flows through the active material. This signal has been experimentally measured via the magneto-optical Kerr effect (MOKE) to be 5x10${^{-5}}$ $\mu{_B}$/atom in Pt, where $\mu{_B}$ is the Bohr magneton [\onlinecite{Stamm17}]. This small signal is nearly five orders of magnitude smaller than the magnetization of Co (1.72 $\mu{_B}$/atom) and was detected using a lock-in based MOKE system capable of detecting EM field polarization angle rotations on the order of nano-radians, after accounting for heating effects. By integrating similar lock-in methods with the proposed plasmonic interferometer geometry, we anticipate a similarly small signal is resolvable. Measuring this signal via magnetoplasmonics would complement existing spin-resolved optical measurements and provide a parallel experimental pathway to characterize the intrinsic properties of spintronic materials. 

The following analysis assumes the spin accumulation in the active layer is resolvable and examines the intrinsic heterostructure properties that influence the signal. In this device, a charge current,$\it{i}{_c}$, through the active layer generates opposing spin populations with a hyperbolic tangent depth dependence between the top and bottom surfaces [\onlinecite{Stamm17}]. For simplicity, we divide the active material into three separate layers representing the two oppositely polarized spin accumulation regions at each surface and a layer that represents the non-magnetic response of the bulk as indicated in Fig. 1b. The spin diffusion length, $\it{l}{_s}$, represents the internal boundary between the polarized and non-magnetic regions. The dielectric tensor in each polarized region consists of the bulk diagonal component, $\epsilon{_{sh}}$, and off-diagonal components, $\epsilon{_{xz}}$ of opposing parity. For an oscillating charge current,$\it{i}{_{ac}}$, a magnetoplasmonic signal $\lvert\Delta k{_{mp}}\rvert = \lvert k{_x}{^{+\it{i}{_{ac}}}}-k{_x}{^{-\it{i}{_{ac}}}}\rvert$ is expected. However, the value of $\lvert k{_x}{^{\pm\it{i}{_{ac}}}}\rvert$ is now directly dependent on both the thickness of the active layer, $\it{d}$, due to the evanescent nature of the SPP within the heterostructure, and the spin diffusion length $\it{l}{_s}$. In a manner similar to the previous device, $\lvert k{_x}{^{\pm\it{i}{_{ac}}}}\rvert$ is obtained by numerically solving a system of equations consisting of the wave equation in each material and the determinant of a matrix formed from enforcing boundary conditions at each interface (Supplementary Information S3) [\onlinecite{Torrado13}]. 

We characterize the SPP field intensity and $ L{_{SPP}}$ of this heterostructure in Figs.~\ref{fig:epsart3}a and b, respectively, as a function the isolating dielectric and spin Hall layer thickness, $\it{b}$ and $\it{d}$, and compare the resulting values to a standard air/Au interface. Fig.~\ref{fig:epsart3}a plots the SPP field profile as a function of depth for five structures: a standard air/Au interface (black), and a stack, Au(15)/MgO(2)/XX($\it{d}$)/MgO(2)/Au (blue), where XX is either Pt (red) or W (blue) of thickness $\it{d}$ = 10 nm (solid) or $\it{d}$ = 40 nm (dashed). In a result similar to the first proposed heterostructure, the $\it{d}$ = 10 nm data indicate that the presence of thinner Pt, W, and MgO layers act only as small perturbations to the SPP environment and do not contribute to a significant deviation of the field profile compared to the air/Au interface. However, to maximize the depth-dependent magnetic signal the spin Hall layer thickness $\it{d}$ must be large compared to the evanescent skin depth of the stack to distinguish between the oppositely polarized spin polarizations along the top and bottom surfaces. For the $\it{d}$ = 40 nm cases for Pt and W in Fig.~\ref{fig:epsart3}a, the SPP field intesity along the top surface in each spin Hall layer is similar to the previous cases, but a significant material-dependent deviation of the SPP field profile is observed as $\it{d}$ increases. While the W case exhibits a field profile that tracks the air/Au interface, the fields in the Pt case are drastically reduced by nearly two orders of magnitude at the bottom surface. This implies the spin signal along the top surface of the Pt may be more isolated and distinguishable compared to W.

The dielectric and spin Hall layer thicknesses, $\it{b}$ and $\it{d}$, also directly influence the magnetoplasmonic sensitivity through their influence on the SPP propagation length $L{_{SPP}}$. Fig.~\ref{fig:epsart3}b quantifies and compares the $L{_{SPP}}$ deviation for a Au(15)/MgO($\it{b}$)/XX($\it{d}$)/MgO($\it{b}$)/Au heterostructure to a Au(15)/MgO($\it{b}$)/Au($\it{d}$)/MgO($\it{b}$)/Au heterostructure. In this plot, heterostructures with Pt, W, and Au spin Hall layers are shown in red, blue, and black, respectively. The $L{_{SPP}}$ is extracted from the model for each material as the spin Hall region thickness $\it{d}$ increases for three fixed thickness values of the isolating dielectric MgO, $\it{b}$ = 0 (solid), 2 (dashed), and 5 nm (dot-dashed). Expectedly, the Au stack with no MgO layer has a constant $L{_{SPP}}$ and serves as a base line to compare the additional layers. The 2 and 5 nm MgO isolation layers introduced into the Au stack reduce the $L{_{SPP}}$ most prominently for small $\it{d}$ thicknesses then increase slightly as they approach asymptotic values which are smaller than the $\it{b}$ = 0 nm stack.

A similar $L{_{SPP}}$ reduction is observed in the Pt and W stacks as a function of the isolating dielectric thickness $\it{b}$, but the general behavior of these stacks as $\it{d}$ increases vary significantly from the Au standard. As the W layer increases the $L{_{SPP}}$ of the stack is further reduced in every variant. However, the Pt stack mirrors the Au stack and increases at larger thicknesses eventually producing $L{_{SPP}}$ values greater than an equivalent Au stack. This behavior is a direct result of the different complex optical dielectric constants in each material. We explore this parameter space more fully for a 40 nm active layer in Fig.~\ref{fig:epsart3}c where the $L{_{SPP}}$ values for a general Au(15)/MgO(2)/XX(40)/MgO(2)/Au stack are plotted as a function of the real and imaginary components of the complex optical dielectric constant from -70 to 0 and 0 to 30, respectively. We indicate the values we have used for Pt, W, and Au on this plot for reference. As the real component of the optical dielectric constant decreases to larger negative values longer propagation lengths are expected resulting in a stronger magnetoplasmonic sensitivity for fixed $\lvert\Delta k{_{mp}}\rvert$.

Extracting unambiguous spin diffusion lengths in thin metals and TIs layers has been experimentally challenging. However, the proposed heterostructure and plasmonic readout provide a system where the spin diffusion length can be measured directly without the presence of additional ferromagnets [\onlinecite{Liu12},\onlinecite{Pai12}] and with results that may be interpreted without estimating optical beam profiles [\onlinecite{Stamm17}]. In Fig.~\ref{fig:epsart3}d, we plot the magnetoplasmonic sensitivity $\lvert\Delta k{_{mp}}\rvert L{_{SPP}}$ in a stack containing Pt as a function of the Pt thickness $\it{d}$ for $\it{l}{_s}$ = 5, 10, and 15 nm (solid, dashed, and dot-dashed, respectively). The magnetic signal $\it{q}$ is scaled by the magnetization ratio between Co and Pt, and is assumed to be $\it{q}$ = (1.0+0.29i)x10${^{-6}}$. We distinguish between stacks with ($d > 2l{_s}$) and without ($d < 2l{_s}$) a non-magnetic middle region. In the cases where $d < 2l{_s}$, we assume an effective spin diffusion length $l{_s}{^{eff}}= d/2$ for the top and bottom layers.

The three plots in Fig.~\ref{fig:epsart3}d share similar behaviors as $\it{d}$ increases. As the top and bottom layers are physically separated the magnetoplasmonic sensitivity increases. As the bottom surface of the spin Hall layer is pushed further into the stack, the spin accumulation signal along the top surface becomes more distinguishable. Fig.~\ref{fig:epsart3}a showed that for large $\it{d}$, the evanescent SPP penetration into the stack is exponentially reduced; this effectively isolates the magnetic signal along the top surface and leads to an asymptotic maxima of the magnetoplasmonic sensitivity. The value of this sensitivity, however, is strongly dependent on the spin diffusion length. As $l{_s}$ increases, a larger cross-section of the spin polarized layer interacts with the SPP giving rise to a stronger signal. The shape and behavior of these plots mirror the depth-dependent behavior measured in Ref. [\onlinecite{Stamm17}] using MOKE and suggest that a series of magnetoplasmonic heterostructures with varying spin Hall layer thickness may produce a similar result. This would confirm the viability and signal sensitivity of this magnetoplasmonic readout scheme and offer it as a complementary experimental technique to measure small magnetic signals along interfaces near the surface.

The strength of the maximum sensitivity in Fig.~\ref{fig:epsart3}d are notably small, primarily driven by the small value of $\it{q}$. However, many factors ultimately influence this value including the laser wavelength which alters the optical dielectric constant of the active layer, the material-dependent spin accumulation strength, and the measurement temperature [\onlinecite{Fan13}]. Given these robust experimental variants, our proposed geometry offers a unique testbed to explore the resolvable limits of magnetoplasmonic modulation. Characterizing and understanding these limits may provide pathways to use magnetoplasmonic techniques to study other planar low signal samples, such as transition metal dichalcogenide monolayers (TMDCs) with large SOC and other 2D materials [\onlinecite{Manzeli17}] or interfacial magnetic defects.

We have shown that integrating electrically isolated metals with large SOC or TIs into magnetoplasmonic heterostructures introduces active spintronic devices which may be compatible with existing plasmonic and photonic architectures. The current-induced generation of spin polarization in these materials can drive non-volatile magnetic reorientations that are detectable using magnetoplasmonics. Furthermore, detecting these small, stable, and reproducible spin polarization signals in samples without a ferromagnet offers an opportunity to explore the sensitivity limits of this technique. Finally, the active layers within these magnetoplasmonic heterostructure testbeds are interchangeable and may produce a variety of behavior that could offer customizable performance benefits for a diverse set of memory, logic, and sensing applications.

%


\end{document}